\documentclass[twocolumn,aps,showpacs,superscriptaddress,amsmath,amsfonts,amssymb,floatfix,prl]{revtex4}
\usepackage{dcolumn}
\usepackage{bm}

\usepackage{graphicx}
\usepackage{amsmath}

\begin{document}

\title{Deformed Gaussian Orthogonal Ensemble description of Small-World networks}
\author{J. X. de Carvalho}\email{josue@pks.mpg.de}
\author{Sarika Jalan}\email{sarika@pks.mpg.de}
\thanks{Currently at : Department of Physics, National University of Singapore,
2 Science Drive 3, Singapore 117542}
\affiliation{Max-Planck-Institut f\"ur Physik komplexer Systeme, 
N\"othnitzer Stra$\beta$e 38, D-01187 Dresden, Germany} 
\author{M. S. Hussein}\email{hussein@if.usp.br}
\affiliation{Instituto de F\'{i}sica, Universidade de S\~{a}o Paulo,
C.P. 66318, 05315-970 S\~{a}o Paulo, S.P., Brazil}

\thanks{Supported in part by the CNPq and FAPESP (Brazil), 
       and the Martin Gutzwiller Fellowship program at the
       Max-Planck-Institut f\"ur Physik komplexer Systeme-Dresden (2007/2008).}

\begin{abstract}
The study of spectral behavior of networks has gained enthusiasm over the last
few years. In particular, Random Matrix Theory (RMT) concepts have proven to be
useful. In discussing transition from regular behavior to fully chaotic
behavior it has been found that an extrapolation formula of the Brody type
can be used. In the present paper we analyze the regular to chaotic behavior of
Small World (SW) networks using an extension of the Gaussian Orthogonal
Ensemble. This RMT ensemble, coined the Deformed Gaussian Orthogonal Ensemble (DGOE), supplies a
natural foundation of the Brody formula. SW networks follow GOE statistics till certain range 
of eigenvalues
correlations depending upon the strength of random connections. 
We show that for these regimes of SW networks where spectral correlations do not follow
GOE beyond certain
range, DGOE statistics models the correlations 
very well. The analysis performed in 
this paper proves the utility of the DGOE in network physics, as much as it has been
useful in other physical systems. 
\end{abstract}

\maketitle

\section{Introduction}

Initiated by two seminal works \cite{SW,BA}, the last decade has
witnessed a spurt in 
activities of network research \cite{SW,rev-Strogatz,rev-network}. Regular and random 
networks are the two limiting cases 
of network topology. For the regular network, each node is connected in a fixed 
pattern to the same number of neighboring nodes; on the other hand, for the random 
network, each node is randomly joined with any other node. Real-world networks show 
the properties which are intermediate of the regular and the random one 
\cite{SW,rev-Strogatz,rev-network}. For example, many real-world networks from diverse 
field have very small diameter but have very high clustering, two characteristics 
shown respectively by random and regular networks. To model randomness and regularity, 
Watts and Strogatz proposed an algorithm to generate popularly known as Small-World (SW) 
network, which has the properties of small diameter and high clustering \cite{SW}. 
Moreover, this model network is very sparse, i.e. network with a very few number of 
edges, which is another property shown by real-world networks. 

The structure of networks is described by its associated adjacency matrix $A$. It is 
defined in the following way: $A_{ij} = 1$ if $i$ and $j$ nodes are connected and zero 
otherwise. We consider only undirected networks. In this case, the adjacency matrix is 
symmetric and consequently has real eigenvalues. These eigenvalues give information 
about some basic topological properties of the underlying network \cite{handbook}.
The fluctuations of these eigenvalues can be studied by Random Matrix Theory (RMT).

There is a long history of applications of random matrix ensembles to model 
fluctuations of the spectra of diverse systems \cite{guhr}. Unfortunately analytical 
results exist only if some ideal conditions are fulfilled by the systems studied. On 
the other hand real physical systems usually depart from these conditions. In order to 
cover these situations other ensembles have been introduced \cite{Dyson}. One such 
class of ensemble is the so-called deformed Gaussian orthogonal ensemble (DGOE) 
\cite{dgoe1,dgoe2,dgoe3}. This ensemble has been proved to be particularly useful 
when one wants to study the breaking of a discrete symmetry in a many-body system such 
as the atomic nucleus. It is also useful for studying transition among classes of ensemble 
such as order-chaos (Poisson $\rightarrow$ GOE) and symmetry violation (2GOE $\rightarrow$ GOE) 
\cite{dgoe4}.
Recently Jalan and Bandyopadhyay show that spectra of various model networks and real 
world networks follow universal random matrix properties 
\cite{SJ_pre2007a,SJ_pre2007b}, intermediate between Poisson and GOE statistics. 
Correlations among eigenvalues of SW networks follow GOE statistics of 
RMT for certain range and after that they deviate from the GOE statistics \cite{SJ_New}.
We believe that the DGOE supplies a RMT basis for the Brody \cite{Brody} distribution 
and gives a more accurate description of the GOE-Poisson transition than the 
Berry-Robnik \cite{Berry} model, which purports to justify the Brody formula from an
RMT stand point. The Brody distribution was used previously in SW statistics
investigation \cite{SJ_pre2007a,SJ_pre2007b}. In the 
present paper we analyze the spectra of SW model networks using DGOE.
Based on the results of reference \cite{SJ_New} we argue, and show 
through numerical simulations that fluctuations of the spectra of the SW model 
follows the description of a transition Poisson-GOE.

\section{Small-World networks}
Watts-Strogatz model of SW network is constructed by rewiring the edges of 
regular ring lattice with probability $p$. This rewiring procedure generates a network 
with some random connections, without altering the number of vertices or edges. For 
$p=0$, structure of the regular lattice or $k$-nearest neighbor coupled network 
remains same; on the other hand, for $p=1$, the regular lattice becomes random 
network. For the intermediate values of $p$, the graph is a SW network: 
highly clustered like a regular graph, yet with small characteristic path length like 
a random graph. This onset of SW property happens for a very small value of 
parameter $p$. Characteristic path length is defined as the number of connections in 
the shortest path between two nodes, averaged over all pairs of nodes. For a network 
of size $N$ and average degree $k$, it scales as $N/k$ if network is regular, and 
$log(N)/log(k)$ if network is random. Clustering coefficient ($C$) is defined as the ratio 
of connections between neighbors to the number of allowed links. For regular graphs 
$C$ is very high ($3/4$), whereas for random graphs it scales as $k/N$. Small-world 
networks show intermediate behavior between these two extremes, with average path 
length being as low as for the random graphs, and clustering coefficient as high as that of 
regular graphs.
This intermediate statistical features of SW networks are reflected in their spectral fluctuations,
and can be nicely described using the DGOE which provides a RMT
basis for the deviation from the GOE behavior of the short range correlation aspect of the
eigenvalues, exemplified through the spacing distribution, and the long
range correlation measured by the $\Delta_3$. In the following we
supply a description of the GOE-Poisson transition within the DGOE.

\section{The Deformed Gaussian Orthogonal Ensemble (DGOE): Transitions Among Universality
         Classes in RMT}
The joint probability distribution of elements  of DGOE has the general
form \cite{dgoe2}
\begin{equation}
	P(H,\alpha,\beta)=Z^{-1}_{N}\exp (-\alpha Tr H^{2}-\beta Tr H_{1}^{2}),
\label{Porto}
\end{equation}
where $Z_{N}$ is a normalization factor and $Tr H$ is the trace of the matrix
$H$. In order to describe two interpolating ensemble the matrix $H$ must 
be chosen as the sum of two terms
\begin{equation}
  H = H_{0} + H_{1},
\end{equation}
where the matrices $H_{0}$ and $H_{1}$ define complementary subspaces of $H$.
According to (\ref{Porto}) for $\beta\rightarrow \infty$ the elements of $H_{1}$ vanish and
$H$ is projected onto the matrix $H_{0}$. Since in this work we are concerned with 
the statistics intermediate between Poisson and GOE we will define $H_{0}$ 
as the Poissonian ensemble. It will be a diagonal matrix with elements given
by
$H_{0,j}=E_{0,i}\delta_{ij}$ whose eigenvalues $E_{0,i}$ are independent random 
variable with Gaussian distribution
\begin{equation}
  \rho_{0}(E) = \left(\frac{\alpha}{\pi}\right)^{1/2}e^{-\alpha E^{2}}
\end{equation}
and variance
\begin{equation}
  \left \langle H_{0ij}^{2}\right\rangle =\frac{\delta_{ij}}{2\alpha}.\label{amor1}
\end{equation}
The elements of the diagonal-less matrix $H_{1}$ are also random
independent variables with zero mean and variance given by
\begin{equation}
  \left \langle H_{1ij}^{2}\right\rangle =\frac{1+\delta_{ij}}{4(\alpha+\beta)}=
              \lambda^{2}\frac{1+\delta_{ij}}{4\alpha},  \label{amor2}
\end{equation}
where $\lambda = (1+\beta/\alpha)^{-1/2}$. When $\beta=0$ ($\lambda=1$)
the ensemble corresponds to the GOE. In the limit $\beta\rightarrow \infty$ 
$(\lambda = 0)$, there will be only diagonal elements and 
the Poisson regime is attained.

The average level density
\begin{equation}
  \rho_{\alpha}(E) = \frac{2}{\lambda\pi}\left(\frac{\alpha}{N}\right)^{1/2}\int_{0}^{\infty}
    \frac{dx}{x}e^{-x^{2}/4\alpha^{2}N}J_{1}(x)
\cos \left( \sqrt{\frac{\alpha}{N}}\frac{Ex}{\lambda}\right) \label{density}
\end{equation}
and the cumulative level density
\begin{equation}
     x_{\alpha}(E)=\frac{2}{\pi}\int_{0}^{\infty}\frac{dx}{x^{2}}
            e^{-x^{2}/4\lambda^{2}N}J_{1}(x)\sin \left( \sqrt{\frac{\alpha}{N}}\frac{Ex}{\lambda}\right),
	\label{unfold}
\end{equation}
were calculated by Bertuola \emph{et al.}  \cite{dgoe5}, who 
observed that formula (\ref{unfold}) provides a more accurate manner 
of unfolding the spectra than the usual polynomial unfolding used in \cite{ueda}. These formulas
work very well in the regime close either to Poisson or GOE
statistics. Intermediate between these statistics there is
a transition regime characterized by a rapid change in statistics from almost Poisson
to almost GOE. In this regime formulas (\ref{density}) and (\ref{unfold}) need corrections
(see \cite{foot}).

\section{Simulations and Results}
Numerical simulations of the SW networks are made by considering ensembles of 
20 networks of size $N=2000$ and average degree $k=20$. The adjacency matrix
was diagonalized numerically and its first and last 300 eigenvalues were
discarded. Since an analytical expression for the average density is still
lacking, the unfolding of eigenvalues was made by fitting the
cumulative density or stair-case function
\begin{equation}
  N(E) = \sum_{i=1}^{N}\Theta(E-E_{i}),
\end{equation}
to Chebyshev polynomial using the linear least squares method.
$E_{i}$ are the eigenvalues of the SW network and $\Theta$ is the unit
step function. 
 
For $p=0$, the corresponding 
adjacency matrix would be a banded matrix with entries {\it one} in the band. As some 
connections are randomized with probability $p$, corresponding adjacency matrix gets 
some non-zero entries outside the band, at the expense of equal numbers of entries of 
$one$ in the band.  The mean value of the elements of these matrices is $p$ and 
variance is $p(1-p)$. Fig.~(1)-(6) plot the adjacency matrix for different rewiring 
probabilities. Left sub-figure of (1) plots the adjacency matrix at the onset of 
SW transition ($p \sim 0.002$). For such a small value of $p$, very few 
connections are rewired and hence adjacency matrix is still almost banded with very 
few connections outside the band. Note that we take average degree of network as 
$k=20$, which leads to a sparse network (i.e. the number of connections is of the 
order of the number of nodes). Left sub-figure of Fig.~(6) plots the adjacency matrix 
for $p=0.2$, for this value of $p$, 20\% of connections are rewired leading to the equal 
number of $one$ outside the band.

Results for the statistics intermediate between Poisson and GOE are obtained by 
diagonalization of an ensemble of random matrices. The mean value of the elements of 
these matrices were taken zero and the variance of the diagonal and off-diagonal 
elements given by (\ref{amor1}) and (\ref{amor2}). The unfolding of the spectra of the 
matrices is done using (\ref{unfold}). In the simulations the values of 
$\alpha=1$ and the size of matrices $N=2000$ are kept fixed. In order to simulate a 
transition Poisson-GOE, ensembles with 100 matrices and different values of 
$\lambda$ are considered. For each value of $\lambda$ we check between the 
density of eigenvalue given by (\ref{density})  and the density of eigenvalues from the 
numerical calculation. If the agreement between the two is poor, the 
simulations are re-run using a corrected version of $\lambda$, and
$\alpha$ (called $\lambda$ and A in \cite{dgoe5})
 \cite{foot}. These corrections are needed especially in
the transitional regime alluded to following the discussion below Eq. (7). 

In the discussion of the deviation of the spacing
distribution from that of Wigner, SW practitioners have used the
Brody distribution,\cite{Brody}, which is given by,
\begin{equation}
P_{\beta}(s) = A s^{\beta}\exp{(-Bs^{\beta+1})}
\end{equation}
where $A$ and $B$ are related to $\beta$ through the normalization
condition. Another distribution which also purports to describe the
transition case was derived by \cite{Berry} using RMT and
semiclassical considerations. The DGOE, which we use in this paper,
supplies a natural RMT for the description of the GOE-Poisson and/or
the Poisson-GOE transitions.

In order to investigate the long-range behavior among the 
eigenvalues of SW model we use the Dyson-Mehta statistics $\Delta_{3}$. It is 
defined as
\begin{equation}
  \Delta_{3}(L;a)=\frac{1}{L}\min_{\substack{B_1,B_2}}\int_{a}^{a+L}dE
            [x(E)-B_1E-B_2 ]^{2} 
\end{equation}
where $B_1$ and $B_2$ are obtained from a least-square fit. $L$ is the average number of spacings 
in the integration interval, and 
$x(E)$ the number of eigenvalues which are less than $E$ (for DGOE $x(E)$ is given 
by (\ref{unfold}) ). $\Delta_{3}$ measures the least-square deviation of the function 
$x(E)$ (the unfolded spectra) from a straight line in the interval $[a,a+L]$.
In order to improve the statistics and avoid the introduction of
correlations we choose successive intervals which overlap by $L/2$ \cite{boh}.
According to RMT, for GOE the expected value for large values of $L$ approaches
\begin{equation}
  \Delta_{3_{GOE}} \sim \frac{1}{\pi^{2}}[\ln L - 0.0687]
\end{equation}
and for Poisson statistics it approaches $L/15$.

In the following we present $\Delta_3$ results for SW networks for various $p$ 
values, and corresponding DGOE. The nearest neighbor spacing distribution of SW networks,
which probes for short range correlations of spectra, for the 
range of $p < p_c$ can be modeled by Brody parameter as described in \cite{SJ_pre2007a}. After 
this values of $p$, which corresponds to the SW transition as defined by 
Strogatz-Watts \cite{SW}, the short range correlations of spectra still follows GOE statistics, but
the long range correlations probed via $\Delta_3$ statistics follows GOE statistics only for certain range, 
and after that deviation from GOE statistics is seen \cite{SJ_pre2007b,SJ_New}. Which indicates 
possible breakdown of GOE theory for SW networks. And hence we turn to the random 
matrix theory of DGOE. Note that for $p > p_c$ the range for which $\Delta_3$ follows GOE 
statistics depends upon
the size and average degree of the network as well \cite{SJ_New}.

\begin{figure}[ht]
\centering
\includegraphics[width=0.45\columnwidth, height=3.5cm]{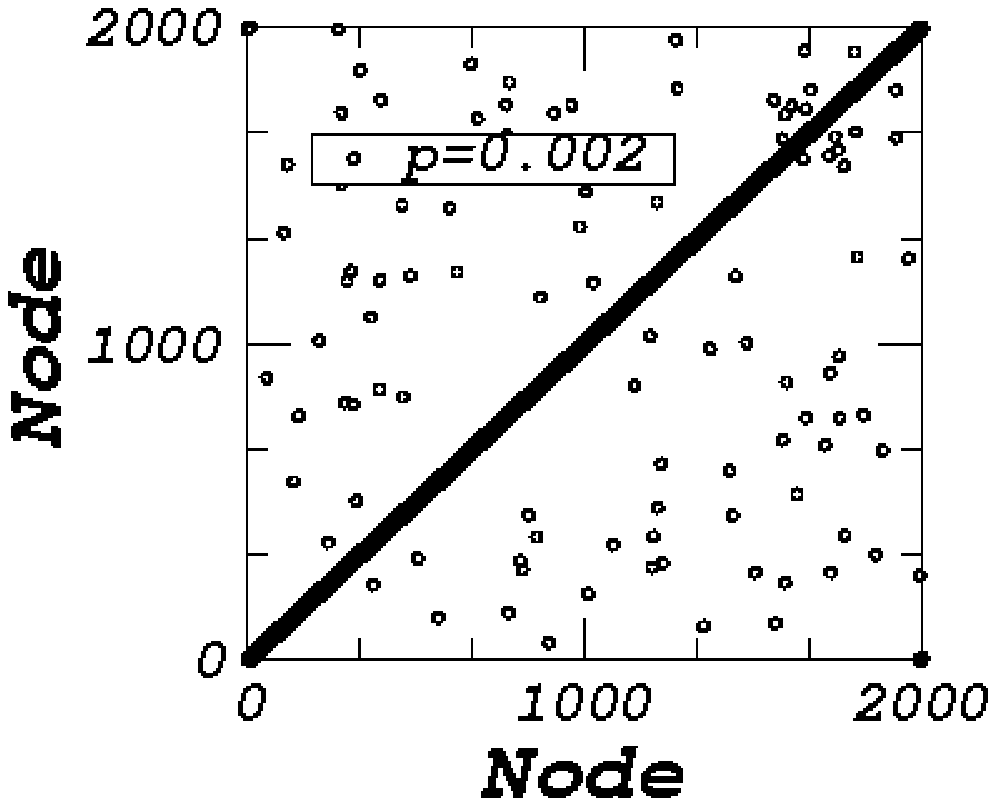} 
\includegraphics[width=0.45\columnwidth, angle=0]{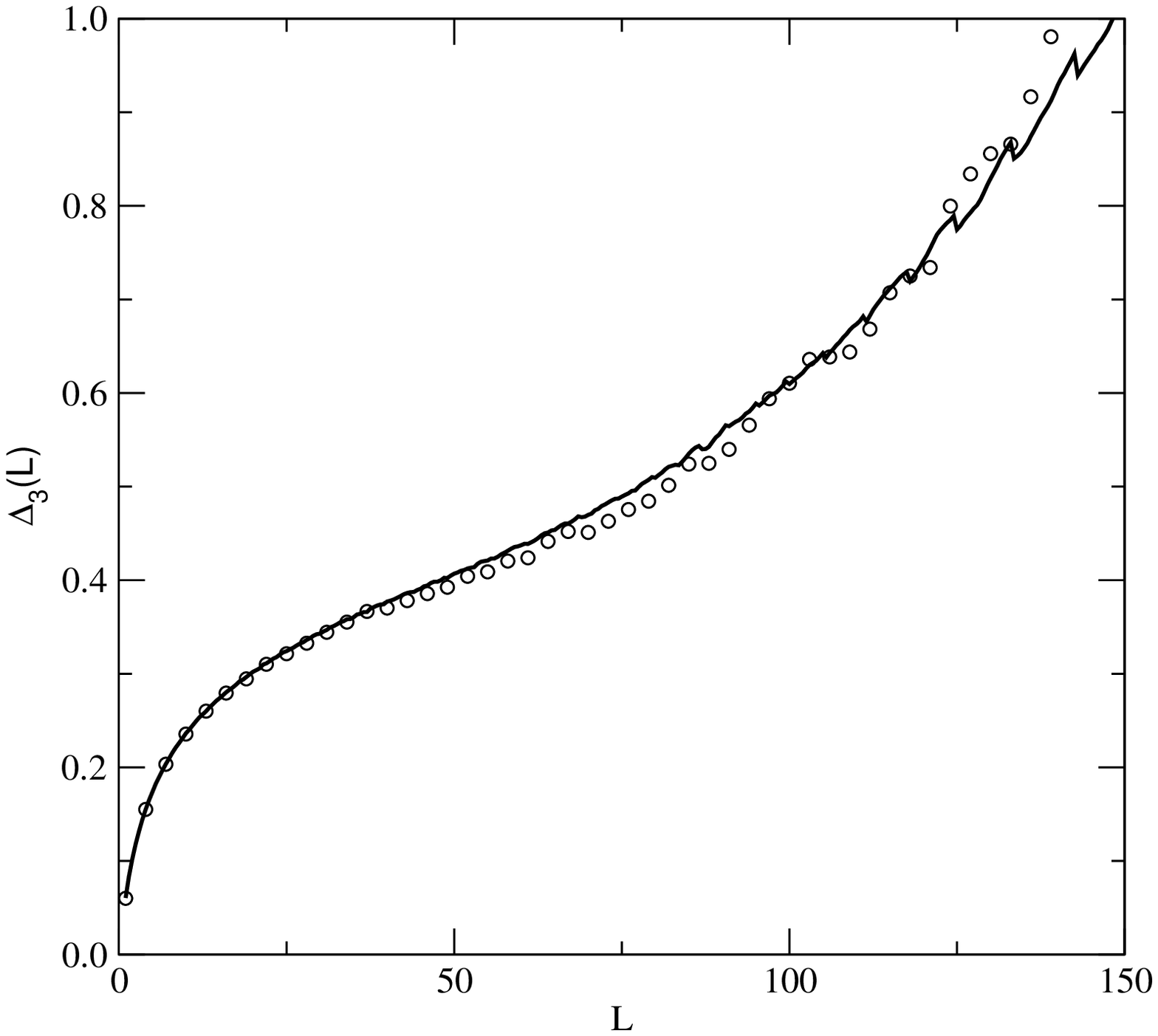}
  \caption{Left sub-figure plots adjacency matrix, and right sub-figure plots the
spectral rigidity as a function of $L$ for the adjacency matrix of SW model with 
$p=0.002$ (circle) and the DGOE with $\lambda =0.0065$ (full line). $\Delta_3$ statistics for 
SW is plotted for 20 sets of random realizations of rewiring.}
           \label{DGOE1}
\end{figure}

\begin{figure}[ht]
  \centering
\includegraphics[width=0.45\columnwidth, height=3.5cm]{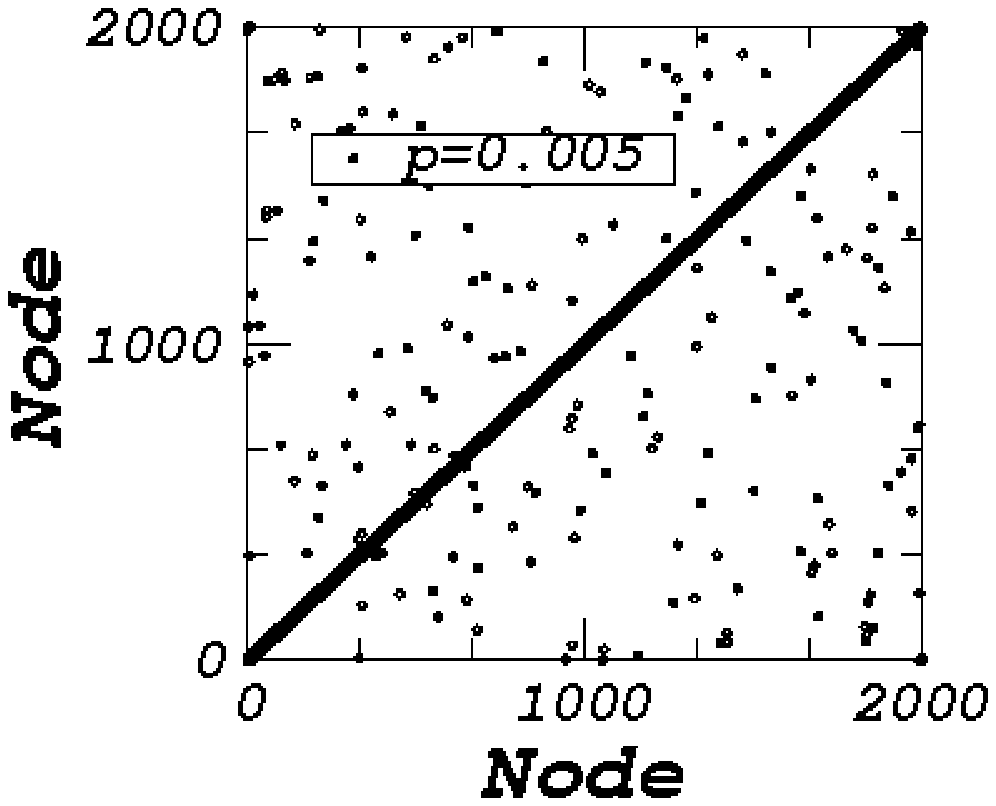} 
   \includegraphics[width=0.45\columnwidth, angle=0]{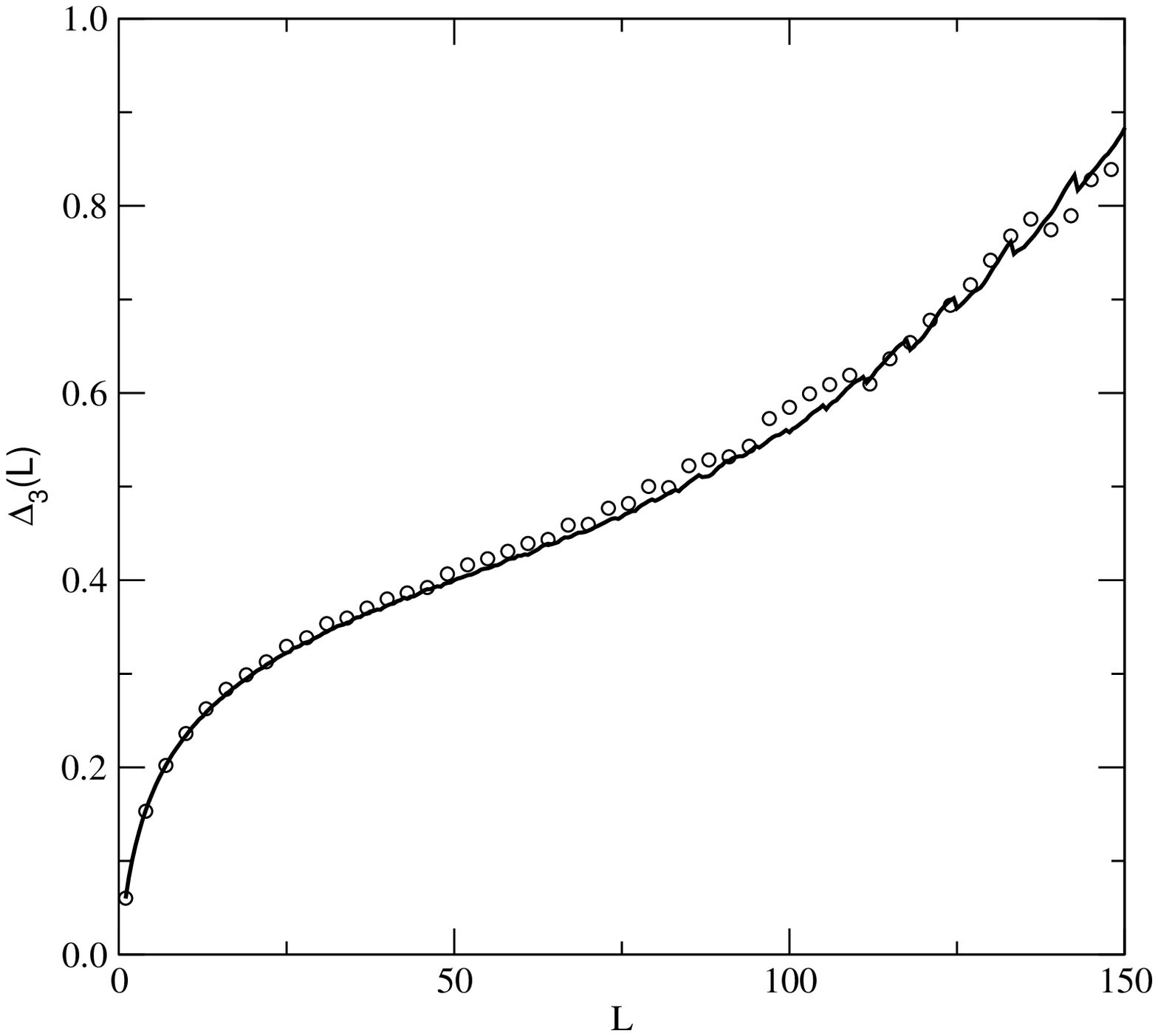}
  \caption{Same as figure 1, for $p=0.005$ and $\lambda =0.0070$.}
           \label{DGOE2}
\end{figure}

\begin{figure}[ht]
  \centering
\includegraphics[width=0.45\columnwidth, height=3.5cm]{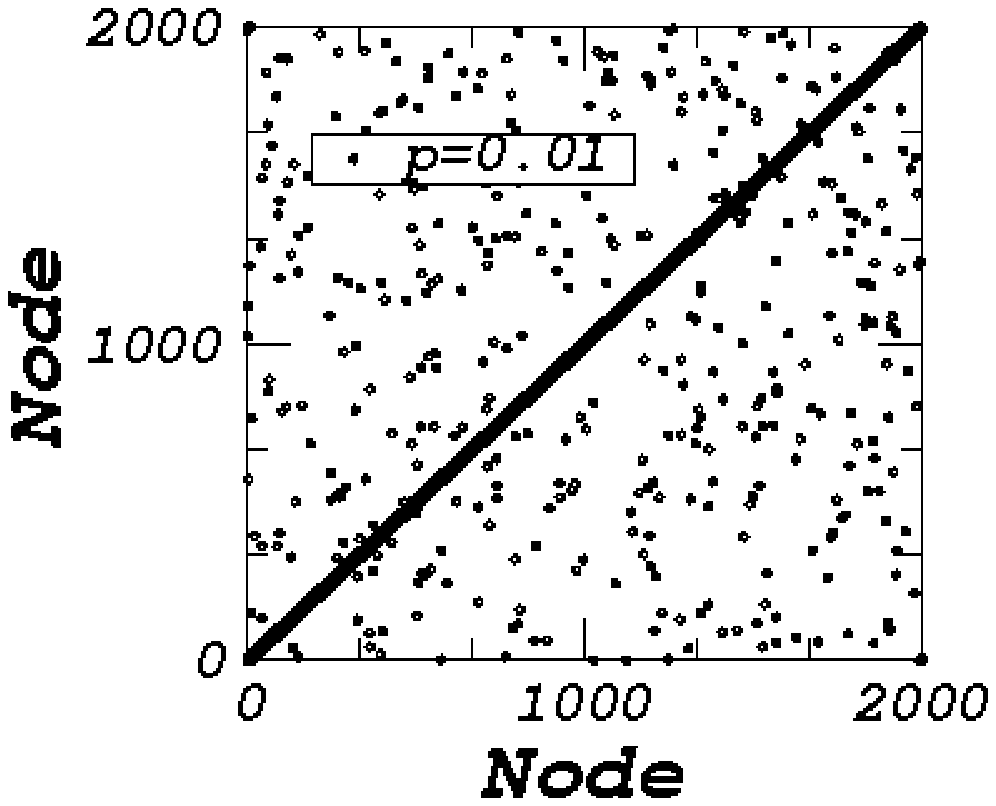} 
   \includegraphics[width=0.45\columnwidth, angle=0]{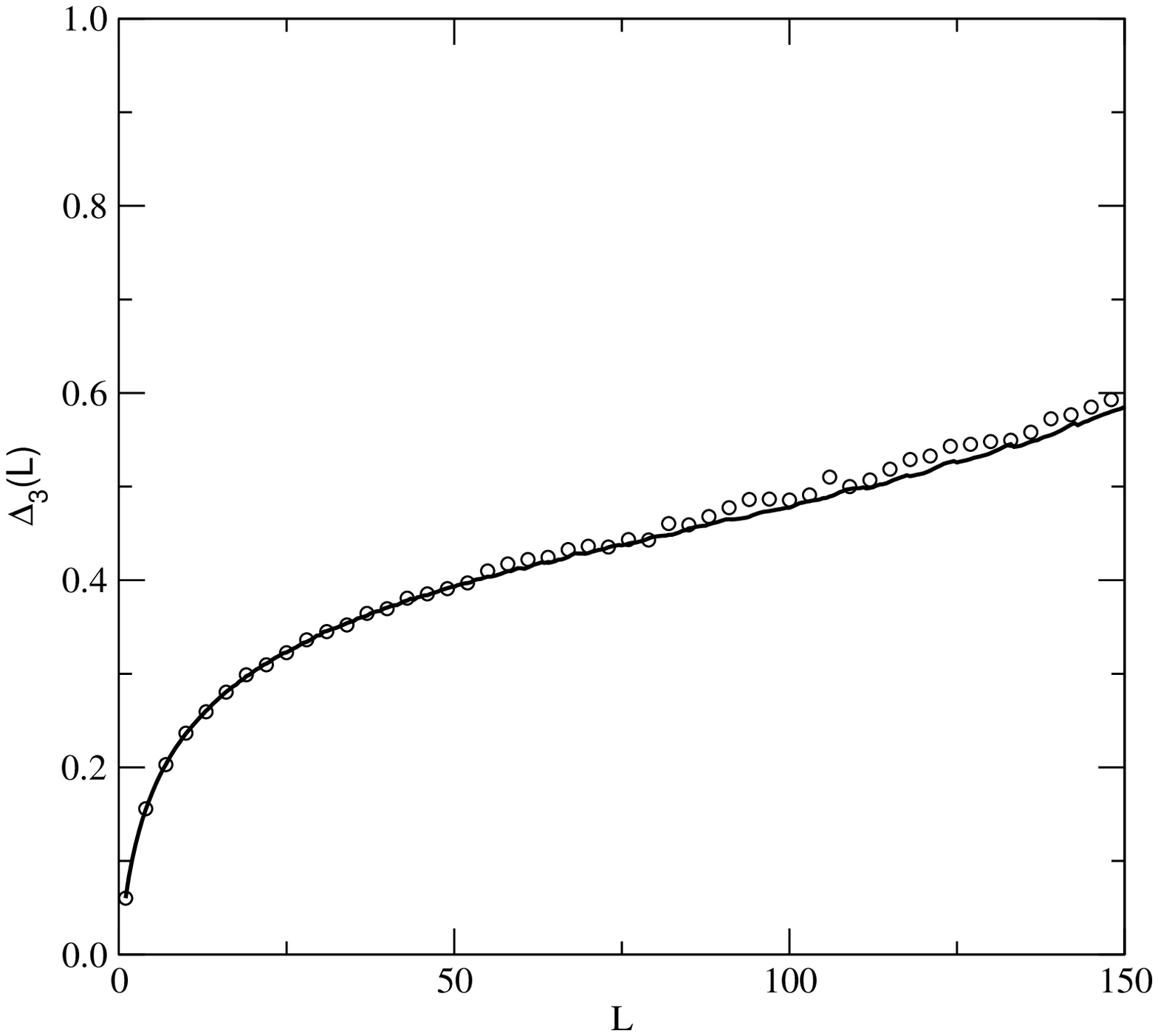}
  \caption{Same as figure 1, for $p=0.010$ and $\lambda =0.010$.}
           \label{DGOE3}
\end{figure}

\begin{figure}[ht]
 \centering
\includegraphics[width=0.45\columnwidth, height=3.5cm]{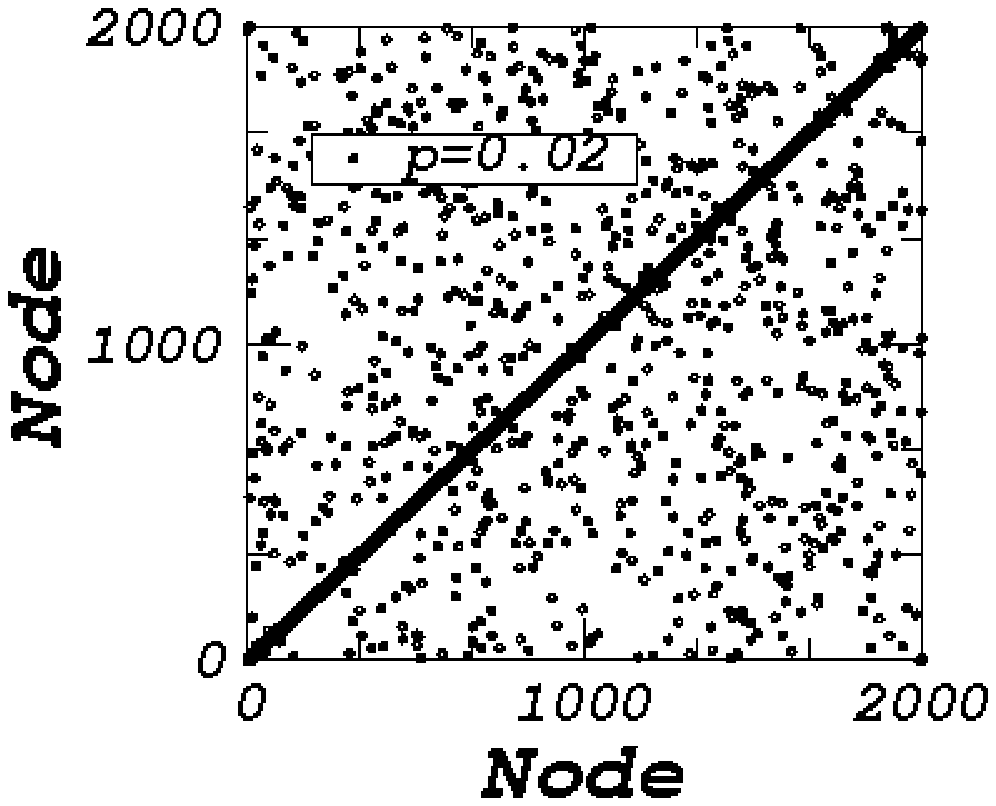} 
    \includegraphics[width=0.45\columnwidth, angle=0]{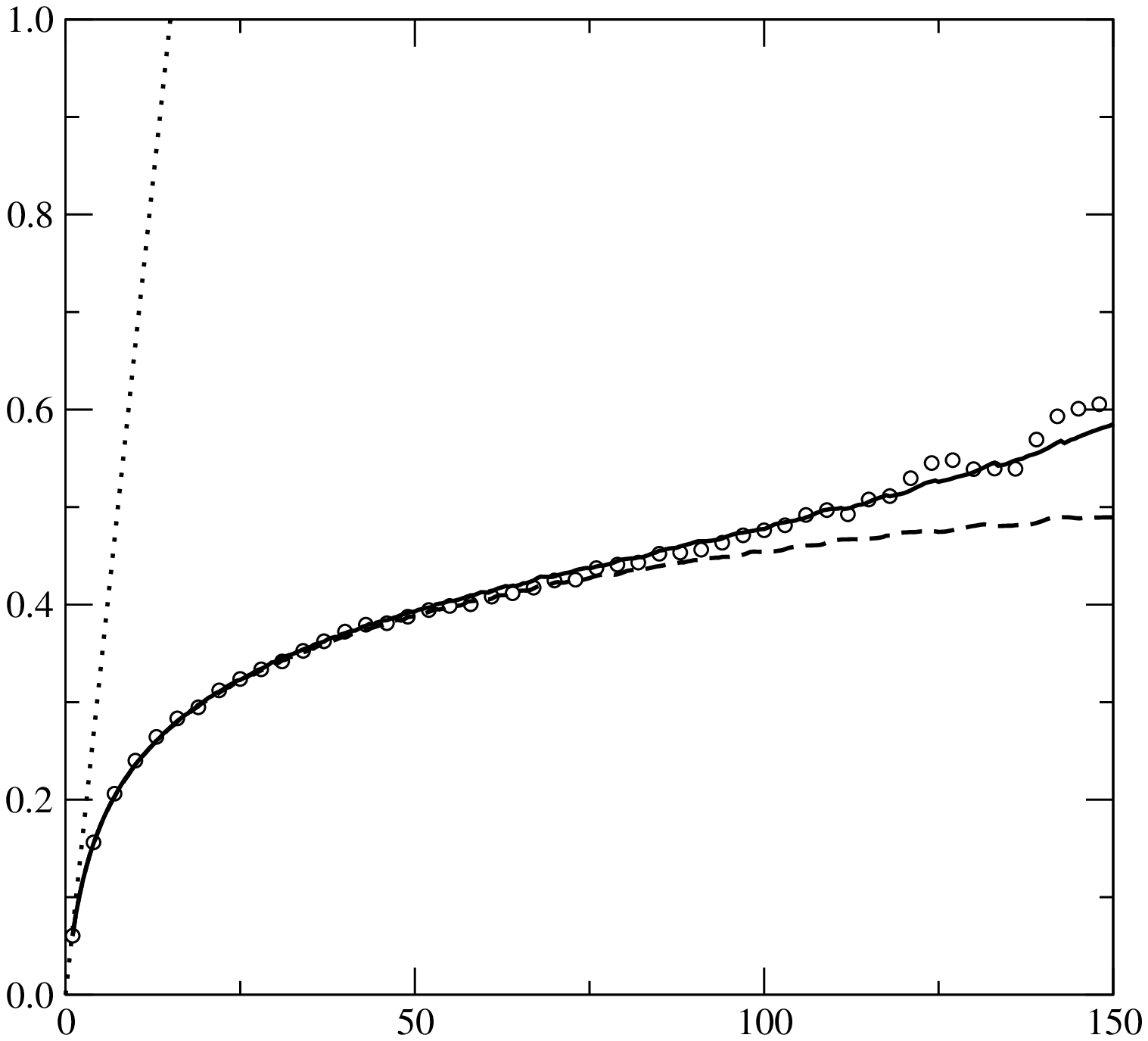}
  \caption{Same as figure 1, for $p=0.020$ and $\lambda =0.010$. Also 
Poisson(dotted) and GOE(dashed) $\Delta_3$ are shown for the comparison.}
           \label{DGOE4}
\end{figure}

\begin{figure}[ht]
 \centering
\includegraphics[width=0.45\columnwidth, height=3.5cm]{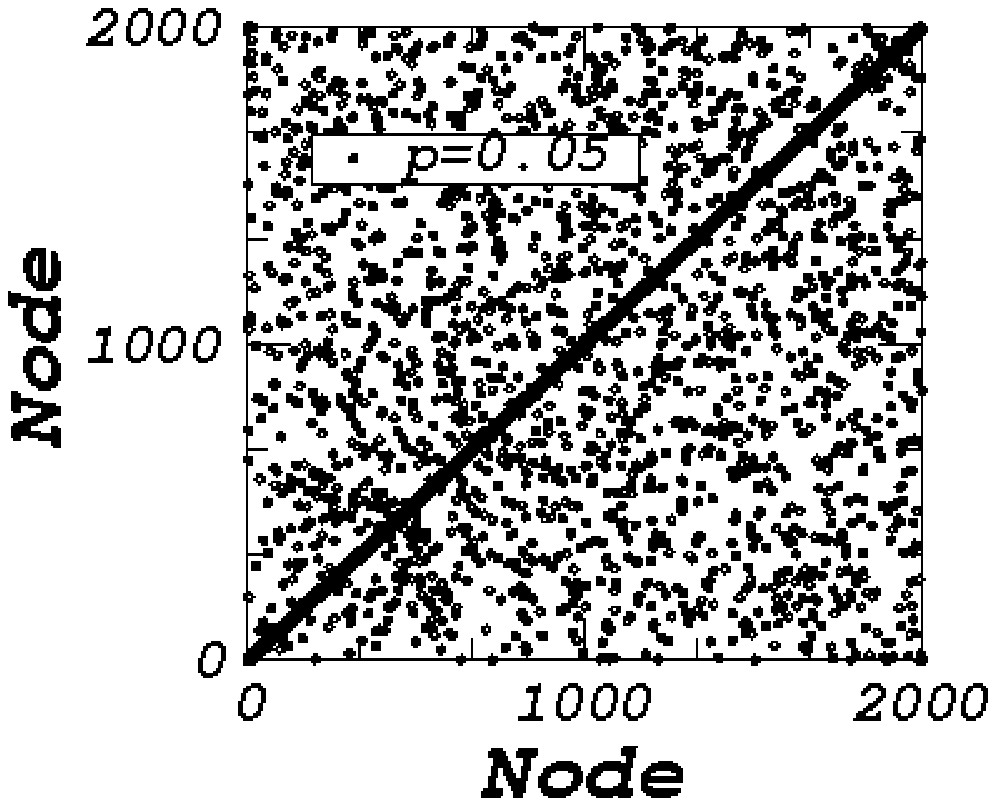} 
    \includegraphics[width=0.45\columnwidth, angle=0]{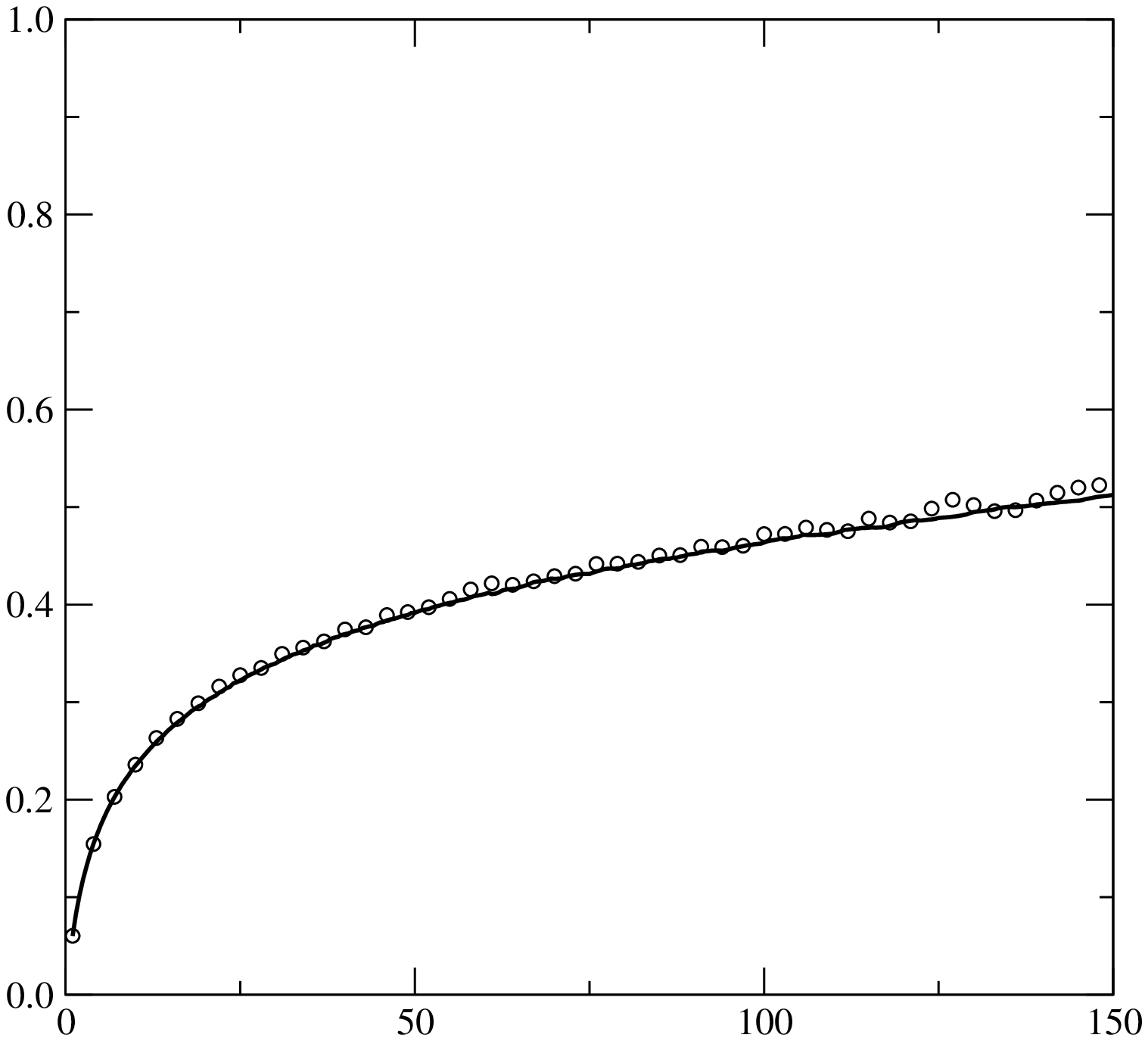}
  \caption{Same as figure 1, for $p=0.050$ and $\lambda =0.015$.}
           \label{DGOE5}
\end{figure}

\begin{figure}[ht]
  \centering
\includegraphics[width=0.45\columnwidth, height=3.5cm]{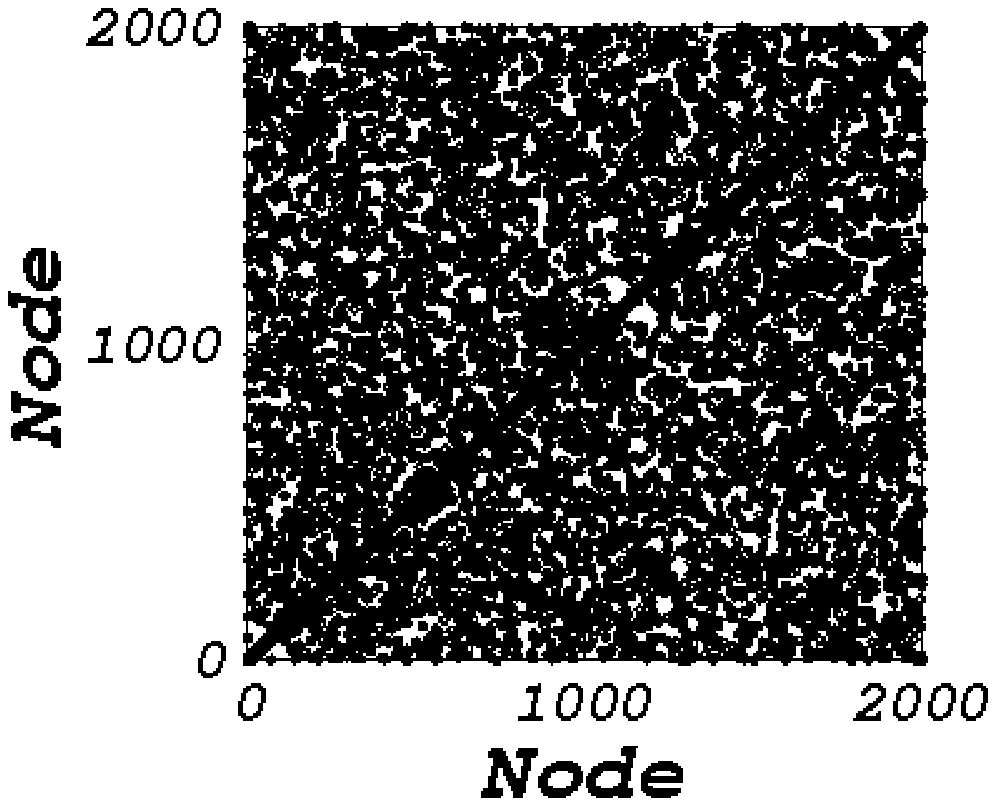} 
    \includegraphics[width=0.45\columnwidth, angle=0]{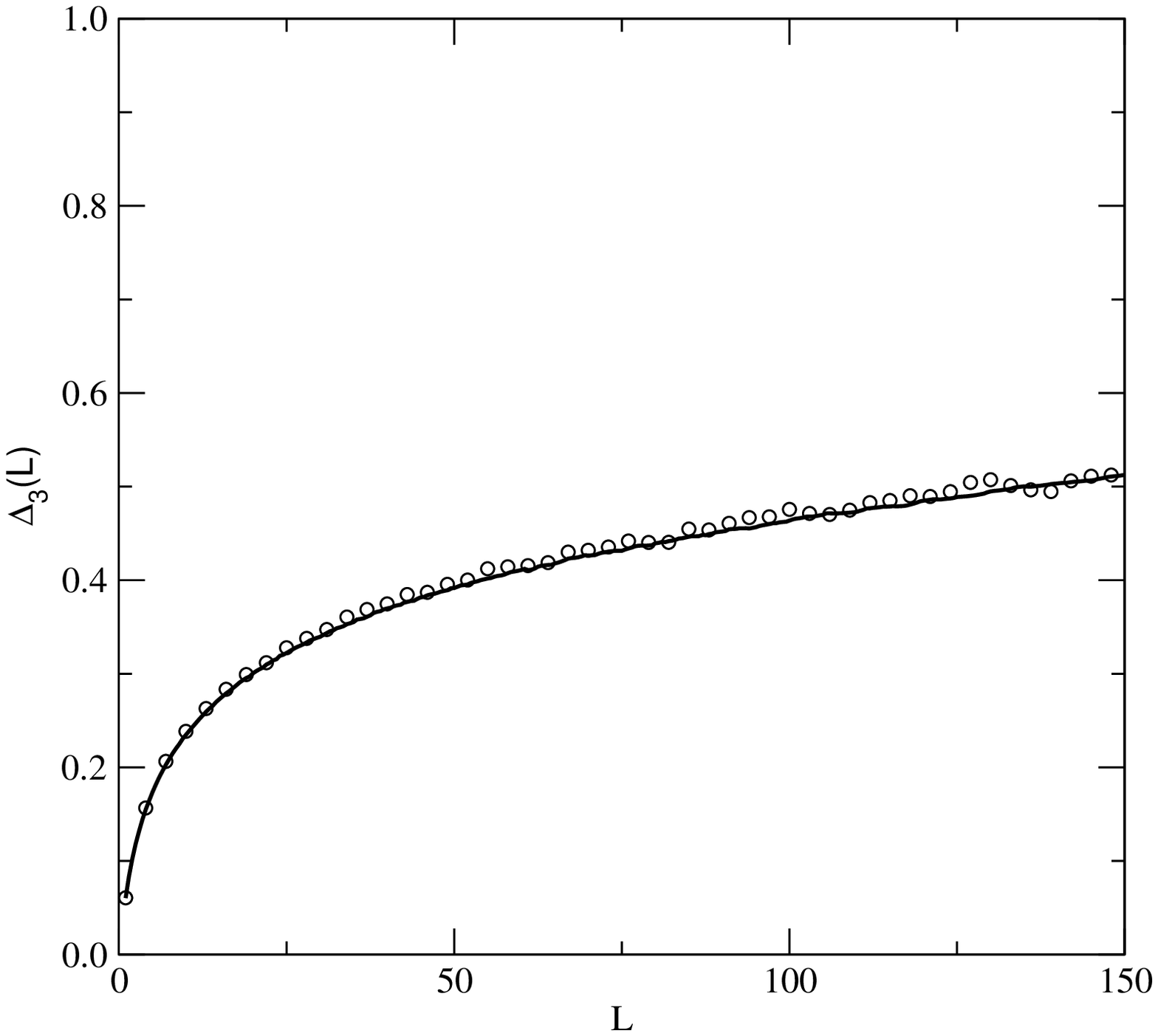}
  \caption{Same as figure 1, for $p=0.20$ and $\lambda =0.015$.}
           \label{DGOE7}
\end{figure}

In Figs.~(1) to (6), the spectral rigidity $\Delta_{3}$ are presented for the different
values of $p$. The values of $p$ varies from $p_{c}=0.002$,
corresponding to the onset of SW behavior,
to $p=1$ which corresponds to a random graph. Each figure also depicts the 
$\Delta_{3}$ for DGOE describing a transition Poisson-GOE. The DGOE
simulations are performed for matrices of size N=2000. Note that for each value 
of $p$ it is possible to find a correspondent $\lambda$ such $\Delta_{3}$ for which DGOE 
fits $\Delta_{3}$ for SW model. The values of parameters $p$ and $\lambda$  
are listed in the table \ref{tablea}. Using the criteria developed in \cite{entropia} we 
find that the critical value of $\Lambda$ which separates the chaotic (random) from 
regular regime is $\Lambda_{c} \approx 0.15 $. Therefore before this value of $p$ the SW 
model is still in the regular regime, although the distribution of nearest neighbor spacing is totally 
compatible with a GOE description. The multi-peaks in the density of 
eigenvalues for these values of $p$ \cite{SJ_New, Farkas} also supports this finding, 
because it indicates that the network still has large amount of regularity. In 
Fig.~(3) to (6) the values for $p$ are increased and the $\Delta_{3}$ comes into the 
chaotic (random) regime.
As the value of $p$ increases the spectra of $SW$ becomes closer to the
GOE prediction. In other words, the local regularity is gradually destroyed and 
the network becomes random. The DGOE description which we are using to 
model SW to random network behavior, shows that for $p \ge 0.05$, behavior 
of $\Delta_3$ statistics can be modeled by a single value of $\lambda=0.015$. It 
suggests that under the framework of DGOE description, the network with $p \sim 0.05$ 
has as much symmetry as for a complete random network ($p \sim 1$).
  

%



\section{Conclusion and Discussion}

According to the RMT the Poisson statistic describes systems with localized states on 
certain bases and uncorrelated spectrum. On the other hand the GOE describes systems 
that become ergodic in the thermodynamic limit and have correlated spectra.
For $p=0$ we have the ring graph which possesses $N$ symmetry (rotational symmetry). The 
numerical calculations of the spectra show several degenerate eigenvalues 
\cite{SJ_New}. There is no level repulsion and the spectra of ring graph should 
follow the Poisson statistics. However, as the value of the parameter $p$ is 
increased gradually the rotational symmetry is destroyed and coupling among the 
eigenstates takes place. The spectra gradually suffer a transition from Poisson 
statistics to GOE. For $p \ge p_{c}$ the spacing distribution, $P(s)$, agrees with GOE 
description, however $\Delta_{3}$ statistics shows some part in the regular regime.
This leads us to conclude that for $p=p_{c}$ the local regular structure is destroyed 
and short-range correlation between eigenvalues is well described by GOE. However 
 some residual local regular structure is still present and the long-range correlation 
among the eigenvalues measured by $\Delta_{3}$ is intermediate
between Poisson and GOE. This residual regular structure is merely
connected to the symmetric nature of the SW ring. This implies a
symmetry constraint in the distribution and the existence of
pseudo-periodic orbits. Such effects leads to a $\Delta_{3}$ which is
a linear combination of a regular, $\frac{15}{L}$, term plus the GOE
term \cite{BP}.  Note that the $P(s)$ is less
sensitive to the finer details of the statistics than
$\Delta_{3}(L)$. The behavior of the SW level statistics (
both $P(s)$ and $\Delta_{3}(L)$) in
this regime can be completely modeled by DGOE which was constructed to
deal with such situations ( Constrained GOE). Finally, for
$p \ge 0.05$ the GOE description of $P(s)$ and $\Delta_{3}(L)$ is recovered.\\

Before ending, we give a detailed assessment of the effect of the size of the
random matrices on the results of the statistical analysis. We have extended our
study above to sizes N = 500, 1000, besides N= 2000. For each case we have performed
the simulations and the subsequent DGOE analysis. Space limitation does not
allow us to present our results in the form of figures but we have collected the
relevant information in the table alluded to above, Table \ref{tablea}. 

\begin{table}[ht]
  \begin{center}

\begin{tabular}{|c|c|c|c|}
\hline
     $p$ &  $N$ & $\lambda$    &   $\Lambda$  \\
\hline
      & 500   & 0.0060 & 0.0180   \\
  0.002 & 1000  & 0.0034 & 0.0116   \\
      & 2000  & 0.0065 & 0.0845   \\
\hline
      & 500   & 0.0090 & 0.0405  \\
  0.005 & 1000  & 0.0050 & 0.0250  \\
      & 2000  & 0.0070 & 0.0980  \\
\hline
      & 500   & 0.0110 & 0.0605  \\
  0.010 & 1000  & 0.0065 & 0.0422  \\
      & 2000  & 0.0100 & 0.2000  \\
\hline
      & 500   & 0.0140 & 0.0980 \\
  0.020 & 1000  & 0.0085&  0.0722 \\
      & 2000  & 0.0100&  0.2000 \\
\hline
      & 500   & 0.0220 & 0.070 \\
  0.050 & 1000  & 0.0120 & 0.144 \\
      & 2000  & 0.0150&  0.450 \\
\hline
      & 500   & 1.0 &  500\\
  0.200 & 1000  & - &  - \\
      & 2000  & 0.0150&  0.45\\
\hline
      & 500   & 1.0 & 500 \\
  1.000 & 1000  & 1.0 & 1000 \\
      & 2000  & 1.0 & 2000 \\
\hline
\end{tabular}
 
 \caption{Results of the DGOE analysis of SW networks using different
   sizes of the random matrices. First column indicates the value of rewiring probability p, 
second column shows size, the third column is the DGOE transition parameter, while
 the fourth column is the modified transition parameter $\Lambda$ =
 N$\lambda^{2}$. See text for details.}
            \label{tablea}
  \end{center}
\end{table}

The first column indicates the value of SW
rewiring probability p which is allowed to vary from very small, 0.002 to the allowed
maximum of 1.00. In the second column indicates the size of matrices. The
last two columns indicate the deduced DGOE parameters $\lambda$ and $\Lambda$
(see the discussion of the DGOE in the section following the Introduction). As a
reminder, the parameter $\lambda$, which takes the values inside the interval
0-1, measures the degree of deviation of the statistics from a pure GOE ( or
pure Poisson). The results shown in table \ref{tablea} clearly indicate that the SW
network is a rigid GOE ensemble, regardless to the size for large values of
p. The size does matter, however, for small values of p, where one sees a clear
dependence of $\lambda$ on the size of the matrices used in the DGOE simulations.

In conclusion, we have performed a statistical analysis of the SW networks
within the DGOE. The analysis clearly
demonstrates the usefulness of the DGOE statistics in supplying a solid basis of an RMT-
based model to describe the chaos-order transitions in such networks. In general terms we conclude that
there is a direct connection between $p$ and $\lambda$, which points
to a natural mapping of SW network onto the DGOE.
Finally, for $p =1 $ when the system is totally random the GOE description is recovered. 

From the random matrix point of view small-world networks
studied here provide a very interesting system where depending upon the rewiring probability one can
see that the short-range and the long-range correlations of the same ensemble of matrices belong to two
different classes of random matrix models. From network point of view the analysis tells that
on the one hand a small amount of random rewiring is enough to introduce short range correlations among
eigenvalues suggesting spreading of randomness in the whole network, on the other hand 
DGOE statistics for long range correlations suggests the nature of symmetry in network. The future
directions of this study is to understand the interplay of dynamical response \cite{syn} 
which is based on the spectra of corresponding adjacency matrix and the symmetries hidden in the 
network under DGOE framework. So far we have only concentrated on the small-world
model network, providing a basis to the DGOE description of networks, future investigations would 
involve studies of real-world networks \cite{realworld}.


\begin{thebibliography}{21}
\bibitem{foot}There are a range of values of $\lambda$ where the formula
(\ref{density}) and (\ref{unfold}), derived in \cite{dgoe5},  must be revised. A way of doing this is
to make the substitution: $\lambda'^{2}\rightarrow \lambda^{2} \{
1+(1+\lambda^{2}/2)[1+\sqrt{1+\lambda^{2}/4}]\} \label{correcao1}$ and
$\alpha'\rightarrow{\alpha[1+\lambda^{2}/2 + \lambda^{2} \sqrt{1+\lambda^{2}/4}]}\label{correcao2}$.
\bibitem{BP}See, however, O. Bohigas and M. P. Pato, Phys. Lett. {\bf B595}, 171
  (2004). In this reference another effect was considered that leads
  to a $\Delta_{3}(L)$ which is a linear sum of a Poisson plus a GOE
  terms. If the statistical analysis is insensitive to, say, $f$
  percentage of the levels, then the long-range correlation measure is
  given by $\Delta_{3}(L)= f \frac{L}{15} +
  (1-f)^{2}\Delta_{3}(\frac{L}{1-f})$. For small values of $f$, which
  is the more likely case ( say, $f = 0.1$), the $P(s)$ is hardly
  affected
by this effect, but the $\Delta_{3}(L)$ is significantly modified,
especially for large $L$, where the dependence is linear.
\end{thebibliography}
\end{document}